\documentstyle[psfig]{mn-ab}

\title[The Millennium Galaxy Catalogue]{The Millennium Galaxy Catalogue: 
Star counts and the Structure of the Galactic Stellar Halo}
\author[Lemon et al.]{D.J.~Lemon$^{1,2}$\thanks{E-mail: djl6@st-andrews.ac.uk}, 
Rosemary F.~G.~Wyse$^{1,3,4}$, J.~Liske$^5$, S.P.~Driver$^2$ and Keith Horne$^1$\\
$^1$School of Physics and Astronomy, University of St Andrews, North Haugh,
St Andrews, Fife, KY16 9SS, UK \\
$^2$ Research School of Astronomy and Astrophysics, Mount Stromlo Observatory, Cotter Road, Weston, ACT 2611, Australia \\
$^3$ Astrophysics, Oxford University, Oxford, UK \\
$^4$ Physics \& Astronomy Dept., The Johns Hopkins University, Baltimore, MD 21218, USA (permanent address) \\
$^5$ Institute for Astronomy, University for Edinburgh, Royal Observatory, Blackford Hill, Edinburgh, EH9 3HJ.}

\date{Accepted
...... Received .....}

\pagerange{\pageref{firstpage}--\pageref{lastpage}}

\begin{document}

\label{firstpage}

\maketitle

\begin{abstract}

	We derive a star catalogue generated from the images taken as
part of the $\sim~37.5~{\rm{deg}}^{2}$ Millennium Galaxy
Catalogue. These data, alone and together with colours gained from the
Sloan Digital Sky Survey Early Data Release, allow the analysis of
faint star counts (B$_{\rm MGC} < 20$) at high Galactic latitude
(41$^{\circ}<b<63^{\circ}$), as a function of Galactic longitude
(239$^{\circ}<l<353^{\circ}$). We focus here on the inner stellar halo,
providing robust limits on the amplitude of substructure and on the
large-scale flattening. In line with previous results, the thick disk,
an old, intermediate-metallicity population, is clearly seen in the
colour-magnitude diagram. We find that the Galactic stellar halo within $\sim 10$~kpc (the bulk of the stellar mass) is
significantly flattened, with an axial ratio of ${\rm (c/a)
=0.56\pm0.01}$, again
consistent with previous results. Our analysis using counts-in-cells,
angular correlation functions and the Lee 2D statistic, confirms  
tidal debris from the Sagittarius dwarf but finds little evidence for other substructure in the inner halo, at heliocentric distances of $\la
5$~kpc. This new quantification of the smoothness in coordinate space
limits the contribution of recent accretion/disruption to the build-up
of the bulk of the stellar halo.

\end{abstract}


\section{Introduction}
The basic stellar components of the Milky Way are the thin disk, thick
disk, stellar halo and central bulge, albeit that the
inter-relationships and distinction amongst different components
remains subject to some debate (e.g.~Norris \& Ryan 1991). Quantifying
the properties of the stellar components of the Milky Way Galaxy is of
wide importance, since the multi-variate stellar distribution function
is a product of Galaxy formation and evolution and in turn constrains
those processes that are important during and after the formation of
the stars. The thick disk of the Milky Way galaxy was introduced by
Gilmore \& Reid (1983) based on their deep star counts towards the
South Galactic Pole that were best fit by including a component with a
scale-height some 3--4 times that of the old thin disk; this component
has the characteristics of `Intermediate Population II' (Oort 1958)
and is clearly seen in earlier star counts (Elvius 1965; Weistrop 1972; Yoshii
1982). That the stellar population of the thick disk is distinct from
that of the halo is seen clearly in colour magnitude diagrams derived
from star count surveys (e.g.~Fig.~2 of Gilmore, Wyse \& Kuijken 1989;
Chen et al.~2001; Fig.~4 here), and many derivations of the field
kinematics and metallicity distributions have established its
existence definitively.

However, even the structural parameters of these major stellar components of 
the Galaxy are less well-established, and deep wide-area star counts are
important in their determination. The flattening of the stellar halo,
when combined with metallicity and kinematic information, can
distinguish between models in which the halo formed with a little, or with a
lot, of gaseous dissipation, and constrains the flattening of the
dark matter halo (e.g.~White 1985). Substructure in phase space is expected in
 hierarchical clustering theories of Galaxy formation, and while the signature
 in kinematics may be more obvious and long-lived (e.g.~Helmi \& White
1999; Helmi et al.~1999), late accretion and merging may produce
observable over-densities in coordinate space (e.g.~Johnston, Hernquist \&
 Bolte 1996; Zhang et al.~2002). The streams from the Sagittarius dwarf 
spheroidal galaxy are extreme examples (e.g.~Yanny et al.~2000; Ibata et al. 
2001; Vivas et al.~2001; Ibata et al. 2002). Placing constraints on the level 
of clustering in coordinate space, for the bulk of the stellar halo, is 
obviously important to constrain recent tidal disruption of, and accretion of 
stars from, satellite stellar systems.

Most previous investigations have utilised small area star counts in a
few selected lines-of-sight (e.g.~Gilmore \& Reid 1983; Bahcall \&
Soneira 1984; Wyse \& Gilmore 1989; Reid \& Majewski 1993).
Investigations into large-scale Galactic structure, and to quantify
the global importance of potentially rare effects -- such as
substructure -- obviously benefit from large-area surveys. While
combination of the data-sets from several independent smaller surveys
is possible (cf.~Reyl\'e \& Robin 2001), the advantages of uniformity
of photometry, star-galaxy classification etc.~all argue for the
superiority of one survey across a wide range of Galactic coordinates.
This has recently become possible through the advent of wide-field CCD
mosaic cameras on medium-sized telescopes, such as that built for the
Sloan Digital Sky Survey (York et al.~2000) and the Wide Field Camera
(WFC) on the Isaac Newton 2.5m Telescope (INT).

In this paper we derive a star catalogue generated from images taken
with the INT/WFC for the $\sim 37.5$~$\rm deg^{2}$ Millennium Galaxy
Catalogue (MGC; Liske et al.~2003). From this catalogue, and in
conjunction with colours gained from the Sloan Digital Sky Survey
Early Data Release (SDSS-EDR; Stoughton et al.~2002), we obtain
reliable star counts, to a limiting magnitude of $B_{\rm MGC} = 20$
mag, as a function of Galactic longitude. The survey is exclusively at
high Galactic latitudes, $b > 40^\circ$, and we investigate the
structure of the Galactic stellar halo, with an emphasis on its
flattening and on the quantification of the statistical significance
of (or lack of) substructure.

	This paper is divided into 5 sections. In Section 2 we describe
the data used in this investigation. In Section 3 and Section 4 we describe the
modeled and observed star counts respectively. The conclusions 
are presented in Section 5.

\section{The Data}

\begin{figure}
\centerline{\psfig{figure=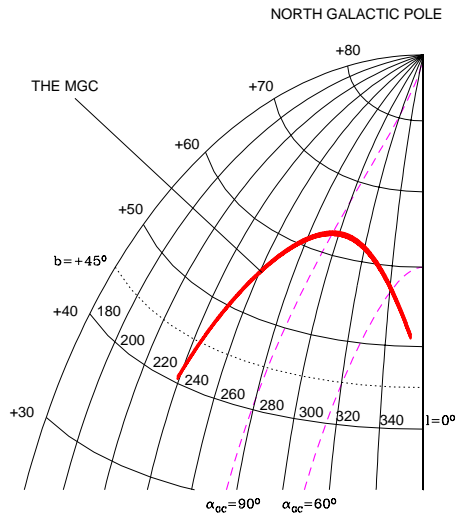,width=\columnwidth}}
\caption[An Aitoff projection of the MGC strip.]{An Aitoff projection
of the MGC strip. The small-dashed line marks the ${\it b}>45^{\circ}$
cut. $\alpha_{GC}$ is the contour of constant angle from the Galactic
center. The long-dashed lines at $\alpha$$_{\rm GC}=90^\circ$ and
60$^{\circ}$ indicate the points at $b$$_{\rm max}$ and
$l$=$340^{\circ}$ respectively.  }
\label{mgclb}
\end{figure}

	Our star catalogue is derived from the imaging survey designed
to provide the Millennium Galaxy Catalogue (MGC; described in detail
in Liske et al.~2003). This consists of a 37.5~deg$^{2}$ single-band
deep ($\mu_{lim}= 26$ mags arcsec$^{-2}$) survey along the equatorial
strip, and overlaps in sky coverage with both the Two Degree Field
Galaxy Redshift Survey (2dFGRS; Colless et al.~2001), and the Sloan
Digital Sky Survey Early Data Release (SDSS-EDR; Stoughton et
al.~2002). All 144 data frames were taken using the Wide Field Camera
on the 2.5m Isaac Newton Telescope situated at La Palma. The WFC is a
prime-focus instrument, and is a mosaic of four 4k$\times$2k thinned
EEV CCDs for the science data, with a smaller 2k$\times$2k Loral CCD
used for auto-guiding. The science CCDs have a pixel scale of
0.333~arcsec~$\rm pixel^{-1}$, and the layout gives a total sky
coverage of $0.29$~deg$^2$ per pointing. Each pointing was observed
for a single 750~s exposure through a Kitt Peak National Observatory
$B$ filter ($B_{\rm KPNO}$). Pointing~1 (field~1) is centered on RA~$=
10^{\rm h} 00^{\rm m} 00^{\rm s}$, DEC~$= 00\degr00\arcmin 00\arcsec$
(J2000) and pointing~$144$ (field~$144$) is centered on RA~$= 14^{\rm
h} 46^{\rm m} 00^{\rm s}$, DEC~$= 00\degr 00\arcmin 00\arcsec$
(J2000); Fig.~\ref{mgclb} displays the MGC strip as an Aitoff
projection in Galactic coordinates; note the emphasis on high Galactic
latitudes.

The MGC astrometry is accurate to $\pm 0.08$ arcsec in both RA and
DEC, and the photometry is internally consistent to $<0.03$ mag (see
Liske et al.~2003). Star-galaxy classification is based on the
`stellaricity' parameter produced by the SExtractor software package,
determined for each object using an artificial neural network that was
trained extensively to differentiate between stars and galaxies (see
Bertin \& Arnouts 1996 for details). All cosmic rays, CCD defects,
satellite trails, diffraction spikes and asteroids have been masked
and removed from the catalogue. As shown in Liske et al.~(2003),
star--galaxy separation is extremely reliable for $16<B_{\rm MGC}<20$;
the star-count investigation in this paper is thus limited to this
magnitude range\footnote{Note that the MGC is complete down to to
$B_{\rm MGC}$=23.5 mag and detects point sources to $B_{MGC} \approx
25$ mag, but accurate star-galaxy separation is limited to an imposed
cut at $B_{\rm MGC}$=20 mag.} (the MGC-BRIGHT catalogue) giving us a
total sample of 42413 stars.

The basic star catalogue is thus based on $B$-band magnitudes. We
first investigate Galactic structure using these single-band data
alone. We then exploit the overlap with the SDSS Early Data Release to
isolate F-stars, thus targeting the turn-off of the stellar halo.

\section{Star count model}

The star count model we use is that developed by Gilmore (1984; see
also Gilmore, Reid \& Hewett 1985 and Wyse \& Gilmore 1989) and
includes a double-exponential thin disk, a double-exponential thick
disk, and a de~Vaucouleurs stellar halo. The thin disk and halo
luminosity functions are based on those of Wielen \cite{wielen}. For
the thick disk the Wielen LF applies only for $M_{V}>4.5$ -- for
$M_{V}<4.5$ the luminosity function of the thick disk follows that of
47~Tuc, the globular cluster with metallicity and age similar to that
of a typical thick disk star (cf.~Gilmore, Wyse \& Jones 1995). Indeed
the colour-magnitude relation of 47~Tuc is adopted for the thick
disk. The colour-magnitude relation of the metal-poor globular cluster
M5 is adopted for the stellar halo.

To facilitate our determination of the flattening of the stellar halo
we held all the structural parameters of this model fixed, with the
exception of the stellar halo axial ratio. We adopted a solar
Galactocentric distance of 8~kpc, an old thin disk scale-height of
325~pc, a thin (and thick) disk scale length of 3.5~kpc, a thick disk
scale-height of 1300~pc and thick-disk normalisation in the mid-plane,
relative to the thin disk, of 2 per cent. The major-axis
de~Vaucouleurs radius for the stellar halo was fixed at 2700~pc. We
made predictions for models with oblate stellar halos, with axial
ratio $(c/a)$ in the range 0.45 - 0.80, at a resolution of 0.01, and
all with a local normalisation of the stellar halo relative to the
thin disk of 0.125 per cent (cf.~Morrison 1993).

With these parameter values, the stellar halo does not contribute
significantly to the star counts until $B \ga 17.5$ mag (shown
graphically in Fig.~\ref{colmag} below). Given our intended aim here
is to investigate the structure of the halo, we will only compare with
data fainter than this limit.

\section{The Observed Star Counts}

\subsection{Constraints on Flattening of the Stellar Halo from $B$-magnitudes Alone}

\begin{figure}
\centerline{\psfig{figure=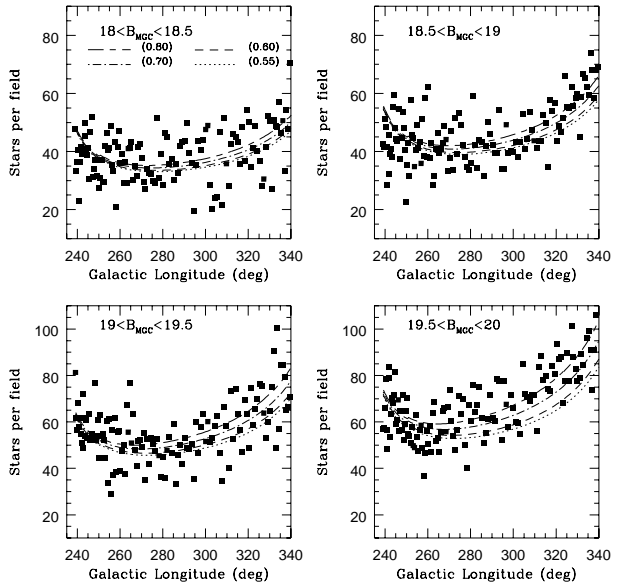,width=\columnwidth}}
\caption{The MGC star counts as a function of Galactic longitude, in
bins of apparent magnitude $B_{\rm MGC}$, for $B_{\rm MGC} > 18$ mag,
compared to the predictions of star count models with a range of
flattenings of the stellar halo. The axial ratio of the stellar halo
is given in parentheses. Note here we have scaled the counts so that
each field covers an area of 0.29 deg$^{2}$.}
\label{smod2}

\end{figure}

\begin{figure}
\centerline{\psfig{figure=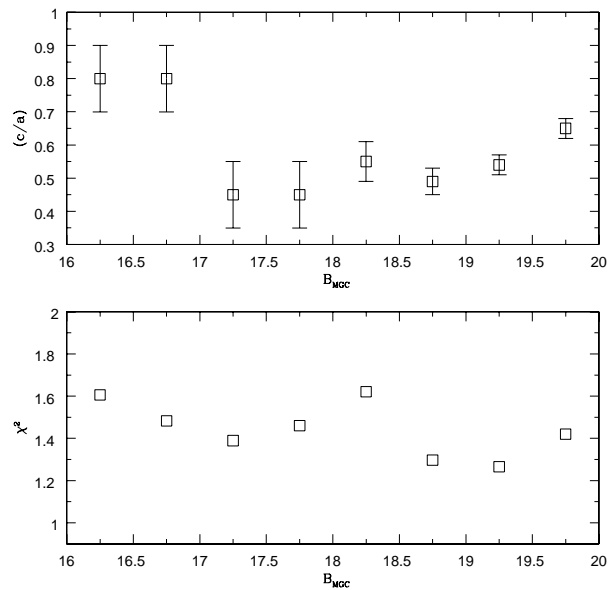,width=\columnwidth}}
\caption{The quality of fit and axial ratio (c/a) of the best-fit
star-count model in each magnitude bin from $16<B_{\rm MGC}<20$. Here
and throughout $\chi^{2}$ refers to the minimum reduced $\chi^{2}$
value. Also, unless otherwise indicated in the figure (through cuts in
$l$ and $b$), for each minimum $\chi^{2}$ estimate there are 144
fields and hence 144 data points (see Fig.~\ref{smod2}) and only one
free parameter (the axial ratio of the stellar halo).}
\label{fig3}
\end{figure}

	Fig.~\ref{smod2} shows how the observed faint star counts,
taken from the MGC-BRIGHT catalogue, vary with Galactic longitude and
magnitude. Also shown are the predicted counts from the model with a
range of flattening of the stellar halo (the value of the axial ratio,
$(c/a)$, is given in parentheses in the Fig.~2, upper left). We
quantified the fits of the models to the data as given in
Table~\ref{alldata} and Fig~\ref{fig3}, assuming that the errors on
the counts are Poisson errors plus a maximum 3 per cent systematic
error to be added in quadrature, derived from our estimates (see
section~4 below) of the reliability of our stellar classification. In
the range $17.0<B_{\rm MGC}<19$ there is a consistent signal of a
flattened halo, with axial ratio (c/a)$\sim 0.5$, with the best fit
giving an axial ratio of (c/a)=0.54$\pm$0.03. In all cases the error
given for the ratio (c/a) is the $\chi^{2}+1$ error ({\it i.e.,}
$1\sigma$). The rise for brighter magnitudes, where there are only $\sim
30$~stars in each field, is significant at only the $\sim 2\sigma$
level. The rise at fainter magnitudes may reflect real changes in the
axial ratio with increasing Galactocentric radius (cf.~Hartwick 1987)
although the signal would have to be produced by intrinsically bright
(and hence distant) tracers. Additionally, the presence of halo
substructure could bias the result (see below, section 4.3).


\subsection{The Colour-Magnitude Diagram}

Colour data for our star catalogue would allow us to identify
particular spectral types of stars. The main sequence turn-off of the
stellar halo is in the F-star region, and isolation of these stars
maximizes the signal from the stellar halo, allowing a more sensitive
testing of the flattening of the halo. Further, we can derive
statistical measures of substructure in the stellar halo from analysis
of the F-star distribution alone.

As noted above, the area covered by the MGC overlaps with part of the
SDSS-EDR, specifically the SDSS run~756, stripe~10. We identified all
the objects in the SDSS-EDR stellar database that lay within the MGC
strip. Each object that was classified as a star in the MGC was then
matched to the SDSS-EDR catalogue using a positional tolerance of
$\Delta \theta$=$\pm$1 arcsec (equivalent to $\sim$3 pixels on the WFC
camera). Of the 42413 MGC stars, 96.7 per cent had SDSS-EDR
counterparts, and there was no case where an MGC star was matched to
more than one SDSS-EDR star. Of the 1396 unmatched MGC stars, 49.79
per cent had $B_{\rm MGC}<15$, all of which fell in the ``flooded
stars region'' of the $B_{\rm MGC}$-Stellaricity plot (see Fig.~10 of
Liske et al.~2003).  This leaves a total of 41718 MGC-stars at $B_{\rm
MGC}> 16$ of which only 701 (2 per cent) have no SDSS-EDR
counter-part. From these 701 unmatched objects, 115 were classified
based on visual inspection by Liske, Lemon \& Driver (Liske et al
2003), and nearly all have $B_{\rm MGC}>19$. Of the remaining 586
stars, 151 fall into the known holes in the SDSS-EDR at $\sim
215.50^{\circ}<{\rm RA}<215.95^{\circ}$ and ${\rm DEC}<-0.21^{\circ}$
and $0^{\circ}<{\rm DEC}<0.21^{\circ}$. A further 24 are clumped in a
strip at $\sim 193.0^{\circ}<{\rm RA}<195.0^{\circ}$ and
$0^{\circ}<{\rm DEC}<0.21^{\circ}$. The rest of the unmatched stars
are evenly distributed across the MGC strip.

\begin{figure}
\centerline{\psfig{figure=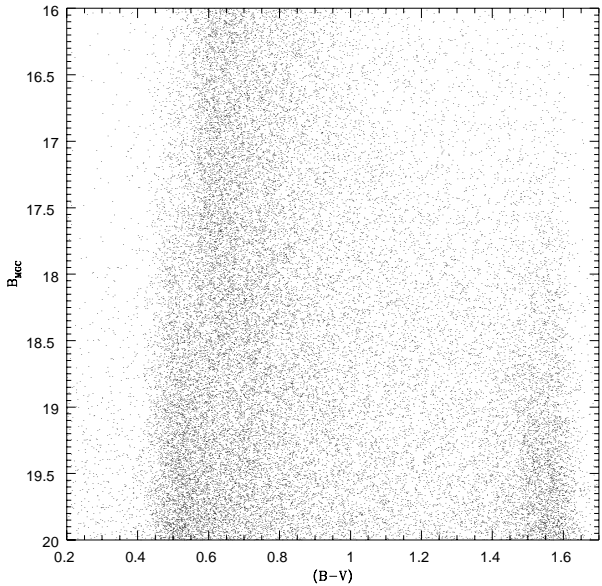,width=\columnwidth}}
\caption{ A CMD plot for the MGC stars in the range $16<B_{\rm MGC}<20$.
}
\label{colmag}
\end{figure}

\begin{figure}
\centerline{\psfig{figure=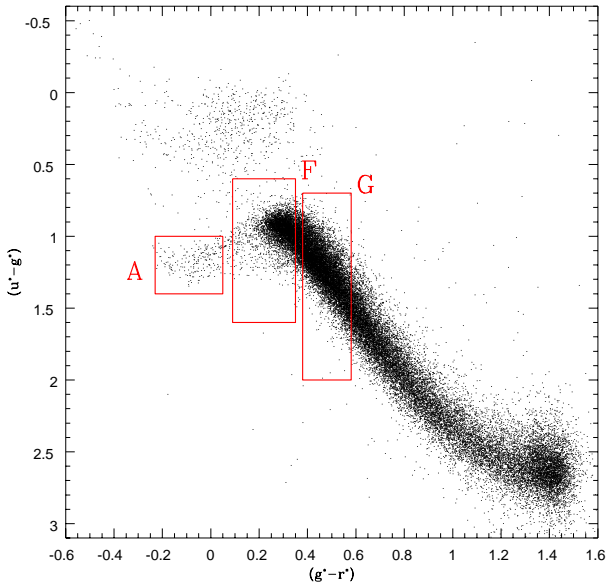,width=\columnwidth}}
\caption{ A colour-colour plot for the MGC stars in the range $16<B_{\rm MGC}<20$. The boxes indicate the selection boundaries for A, F \& G stars. Objects making up the fuzzy patch to the top left are most likely quasars. }
\label{colcol}
\end{figure}

	The SDSS-EDR magnitudes used in this paper are SDSS PSF
magnitudes (see Lupton et al.~2003). Using the 41017 matched
MGC-SDSS-EDR stars we are able to plot a colour-magnitude diagram
(hereafter CMD; Fig.~\ref{colmag}), and a colour-colour diagram
(Fig.~\ref{colcol}), for MGC stars in the range $16<B_{\rm
MGC}<20$. The $(B-V)$ colours are gained via the colour transformation
found in Fukugita et al.~(1996):

\begin{equation}
(g^{*}-r^{*})=1.05(B-V)-0.23
\end{equation}

Three distinct concentrations of stars are obvious in the
CMD of faint stars in high-latitude fields, representing the stellar
halo, the thick disk and the thin disk, as shown in Gilmore \& Wyse
(1985; their Fig.~3) and in Gilmore et al. (1989; their
Fig.~2). The first, seen here in Fig.~\ref{colmag} at
($B-V$)$\sim$0.45, $B_{\rm MGC}>18$, reflects the main-sequence
turnoff colour of metal-poor, old stars, and contains stars that are
members of the stellar halo. The well-defined blue limit suggests a
uniform old age for stars in the halo, with no significant
intermediate-age population (see Unavane, Wyse \& Gilmore 1996 for
quantification of this point, and discussion of the implications for
late accretion into the halo). The second concentration, at
($B-V$)$\sim$0.65 and $B_{\rm MGC}<18$, marks the main-sequence
turnoff of more metal-rich old stars, which reside in the thick disk
(cf.~Gilmore et al. 1985). The final concentration, at
$(B-V)>1.5$, is due to thin disk stars, and reflects the
insensitivity of the B and V bands to cool stars on the main
sequence (see Gilmore \& Wyse 1987 for a comparison of the CMD in
different band-passes). The CMD for the full SDSS-EDR data, in the
SDSS band-passes, was presented by Chen et al.~(2001) where the same features are seen.

It is clear from the morphology of the CMD that one cannot neglect the thick 
disk. It is also clear that the contribution of the stellar halo to our star
 counts is maximised by considering only F-stars, $B-V \la 0.6$, and 
restricting the analysis to stars fainter than $B_{{\rm MGC}} = 18$. 

The colour-colour diagram of the MGC stars, using the SDSS photometry,
is shown in Fig.~\ref{colcol} and provides a basis for spectral-type
selection. Our chosen boundaries for various spectral classes of stars
are as indicated. In principle F-stars cover a wide range in
($g^{*}-r^{*}$) and ($u^{*}-g^{*}$), but after taking into
consideration the sharp halo turn-off at ($B-V$)=0.45 (see
Fig.~\ref{colmag} above), we adopted the more conservative limits
$0.1<(g^{*}-r^{*})<0.3$ and $0.7<(u^{*}-g^{*})<1.0$ (see also Yanny et
al.  ~2001, and Newberg et al.~2002 for similar selection
criteria). The objects making up the fuzzy patch in the top left are
likely to be quasars and contribute 2.5 per cent to the total MGC-SDSS
matched sample. These should be distributed isotropically across the
sky and so while increasing the background somewhat, should not
contribute a false clustering or flattening signal.

\subsection{Substructure in the Halo Star Counts} 

Hierarchical clustering scenarios of galaxy formation, such as in a
Cold-Dark-Matter-dominated Universe, predict that the Milky Way formed
from the merging of many smaller progenitors (see e.g.~reviews of Silk
\& Wyse 1993 and White 1996). The satellite galaxies of the Milky Way
may be surviving examples of the earlier systems. It has been
suggested that a significant part of the {\it outer\/} halo could be
the result of accretion and merging of substructure (cf.~Searle \&
Zinn 1978), with this process on-going, as evidenced by the
Sagittarius dwarf spheroidal (Ibata, Gilmore \& Irwin 1994; Ibata et
al.~1997). The kinematic signature of `moving groups' is long-lived,
producing observable effects at the solar Galactocentric distance and
beyond, even after $\sim 10$~Gyr (Helmi \& White 1999; Johnston et
al.~1999; Helmi et al.~1999; Gilmore, Wyse \& Norris 2002). The
signature of clustering in coordinate space is less robust and decays
faster in time with orbit wrapping, but has clearly been detected for
the tidal streams from the Sagittarius dwarf spheroidal galaxy (Yanny
et al.~2000; Vivas et al.~2001; Ibata et al.~2001, 2002). Newberg et
al.~(2002) identified several over-dense regions in their wide-area
faint F-star sample using SDSS imaging data, several most likely to be
associated with streams from the Sagittarius dwarf, but others that
may be independent structures in the outer halo or disk. Indeed there may be a `ring' around the Milky Way (Yanny et al.~2003; Ibata  et al.~2003).  These identified
over-densities in the SDSS data are at faint magnitudes, $g^\ast\sim V
> 19.4$, corresponding to $B_{\rm MGC} \ga 20$, or heliocentric
distances of $\ga 10$kpc for metal-poor F-stars. Most of the mass of
the stellar halo lies interior to these distances and is sampled by
the brighter stars. Quantification of the level of substructure in
these brighter stars (the inner halo) has heretofore been lacking in analyses of star
counts.

Our present sample is ideal to quantify the clustering in the inner
stellar halo, which contains the bulk of the steellar mass. This
quantification of clustering has two applications, the first to
constrain recent accretion, the second to estimate the possible effect
of small-scale non-uniformities on our derivation of the larger-scale
structure of the stellar halo. Simulations (e.g. Johnston et al. 1996)
have shown that over-densities in coordinate space can survive for
several orbital periods, albeit the alignment is dependent on the
shape of the Galactic potential (longest for spherical potentials) and
on the time dependence of the potential (e.g.~Zhao et al.~1999). The
present analysis may be expected to constrain accretion into the inner  
halo over the last few Gyr, that being several orbital periods
within several kpc of the Sun's location. While the uniform old age of
the bulk of the halo, seen again here in the location of the
well-defined main sequence turn-off in Fig.~\ref{colmag}, argues
against the accretion of systems containing intermediate-age
populations, such as the typical satellite galaxies, in the last $\sim
10$~Gyr being important in general (Unavane et al. 1996), the
accretion of systems with stellar populations similar to the old,
metal-poor halo is better constrained by signatures in kinematics and
coordinate space.

\subsubsection{Counts-in-Cells}

\begin{figure}
\centerline{\psfig{figure=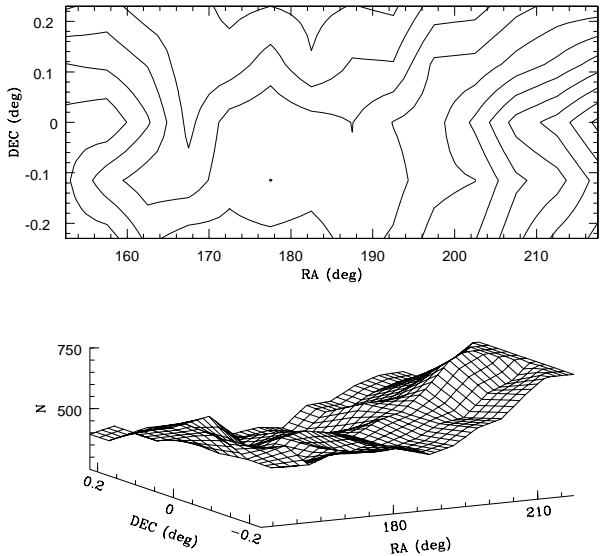,width=\columnwidth}}
\caption{A contour (upper) and surface (lower) plot of the full
stellar distribution. The large scale (galactic) stellar distribution
is clearly visible. The contours are evenly spaced at intervals of
$\Delta N$=50 between $N$=300 and $N$=900.}
\label{posplot}
\end{figure}

\begin{figure}
\centerline{\psfig{figure=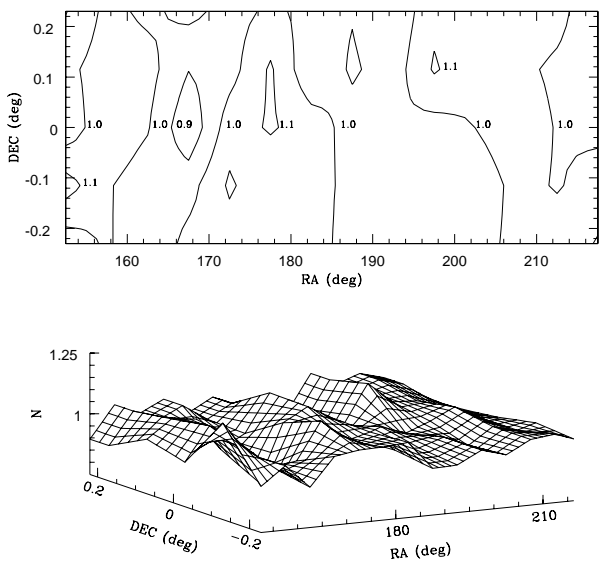,width=\columnwidth}}
\caption{A contour (upper) and surface (lower) plot of the full
``flat-fielded'' stellar distribution with the large scale stellar
distribution removed. The contours are evenly spaced at the levels of
$\frac{N}{N_{\rm model}}$=0.9, 1.0 and 1.1}
\label{posplot_ff}
\end{figure}

A first impression of the clustering of the stars can be seen by a
simple analysis of counts-in-cells. The contour and surface plots
that results from counts in cells of size 0.1~degree in DEC and
5~degree in RA of the full MGC stellar distribution (irrespective of colour, 
and with limiting magnitude $B_{MGC} = 20$) is shown in Fig.~\ref{posplot} 
(note that there are around 10$^3$ stars per square degree at these 
magnitudes, with fields at high latitude and intermediate longitude,
corresponding to the counts per cell here being around 500, and increasing
towards the Galactic Center). The large scale gradient in the stellar
distribution towards the direction of the Galactic Center is clearly
seen. We removed this gradient by fitting a smooth quadratic to the stellar
distribution in RA for each DEC bin, and then dividing the actual
number of stars in each cell by the ``model'' number of stars for that
cell. In this way the large scale stellar distribution is taken out
and only fluctuations due to random noise and/or stellar clustering
are left, as shown in Fig.~\ref{posplot_ff}. Some fluctuations are
seen in this plot, but all are at less than 1$\sigma$ ($\sigma =$0.058) away
from the mean $\left(\overline{\frac{N}{N_{\rm model}}} = 1 \right)$.

\begin{figure}
\centerline{\psfig{figure=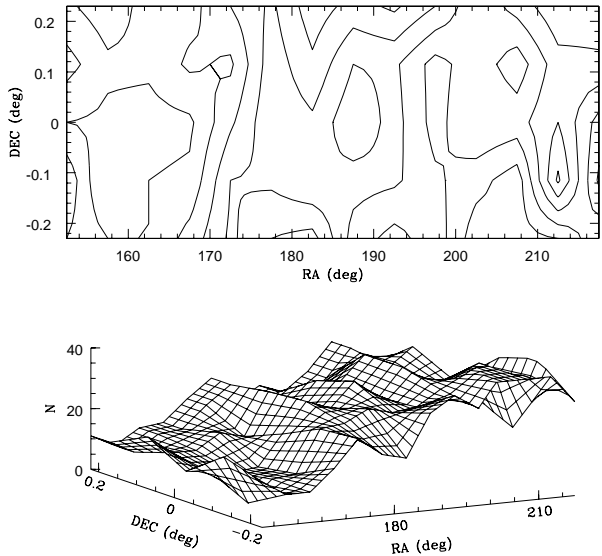,width=\columnwidth}}
\caption{A contour (upper) and surface (lower) plot of the F-star distribution. The large scale (galactic) stellar distribution is clearly visible. The contours are evenly spaced at intervals of $\Delta N_{F}$=5 between $N_{F}$=10 and $N_{F}$=60.}
\label{posplot_f}
\end{figure}

Utilising colour information from the SDSS-EDR and isolating just the
selected F-stars provides the contour and surface plots shown in
Fig.~\ref{posplot_f}; note that by restricting the colour range the
number of stars per field has reduced to several tens rather than the
several hundreds for all stars. Again fluctuations are seen,
superimposed on the large scale gradient, but again at only low level
of significance, at typical level of 1.27$\sigma$ away from the mean.

\begin{figure}
\centerline{\psfig{figure=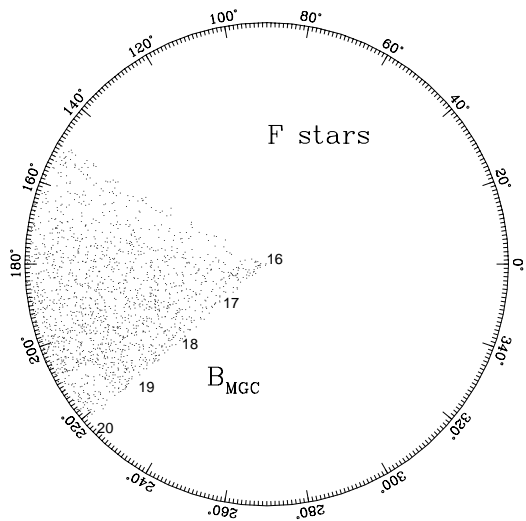,width=\columnwidth}}
\caption{ A `pie-diagram' plot of RA and $B_{\rm MGC}$ for the F-star sample contained within the whole MGC $b$ \& $l$ range.}
\label{rstarplot}
\end{figure}

\begin{figure}
\centerline{\psfig{figure=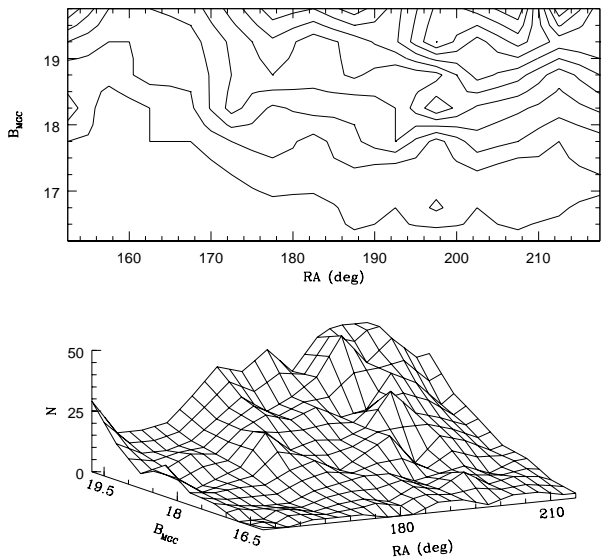,width=\columnwidth}}
\caption{ A contour and surface plot of the F-stars in RA-magnitude space. The contours are evenly spaced at intervals of $\Delta N$=5 between $N$=10 and $N$=60.}
\label{magposplot}
\end{figure}

The colour selection to narrow the spectral type also allows us to use
apparent magnitude as an approximate distance modulus, since the
selected F-stars should have a fairly narrow range in absolute
magnitude. One can then look for clustering along the line-of-sight.
The `pie-diagram' plot of RA versus apparent magnitude (remember the
survey is a narrow strip with little range in DEC) is shown in
Fig.~\ref{rstarplot}, with Fig.~\ref{magposplot} showing the contour
and surface plots in this RA-$B_{\rm MGC}$ plane. The rise in
Fig.~\ref{magposplot} shows the combination of the large scale stellar
gradient and the stellar number count gradient, {\it i.e.,} more stars
exist at fainter apparent magnitudes.

The strongest signal seen in our sample is again at the faintest
magnitudes, B$_{\rm MGC} \ga 19$, and at larger RA$\ga 210^\circ$; it
is difficult to quantify the amplitude of the fluctuation due to the
underlying increase in the star counts in both RA and apparent
magnitude. As we will also see below, these values are close to the
coordinates and apparent magnitude range of the detection of the tidal
arm from the Sagittarius dwarf spheroidal galaxy in A-stars, at
distances of $\sim 40$~kpc; this structure was identified in F-stars
by Newberg et al.~(2002), at apparent magnitudes of $g^\ast\sim 22.5$.
That tidal feature represents material lost in the most pericentre
passage of the Sagittarius dwarf, but tidal streams that were removed
1--2 pericentre passages ago are predicted in some models of the
evolution of the Sagittarius dwarf (e.g.~Johnston et al.~1999) to be
seen in this direction at only 5--10~kpc from the Sun. The predictions
are illustrated in Figure 4 of Kundu et al.~(2002), who made a
preliminary detection of this structure in bright K giant stars.  The
analysis of the 2MASS star counts by Majewski et al.~(2003) also
provides evidence of tidal debris from the Sagittarius dwarf in this
location.  While it is almost inevitable that there will be some
contamination of the F-star data by stars of other spectral types,
this should be relatively small since our colour cuts were
conservatively chosen to minimise any contamination, and thus only
random errors in colour should contribute. Thus we tentatively
identify the apparent upwards fluctuation in counts with a tidal feature from
the Sagittarius dwarf.

\begin{figure}
\centerline{\psfig{figure=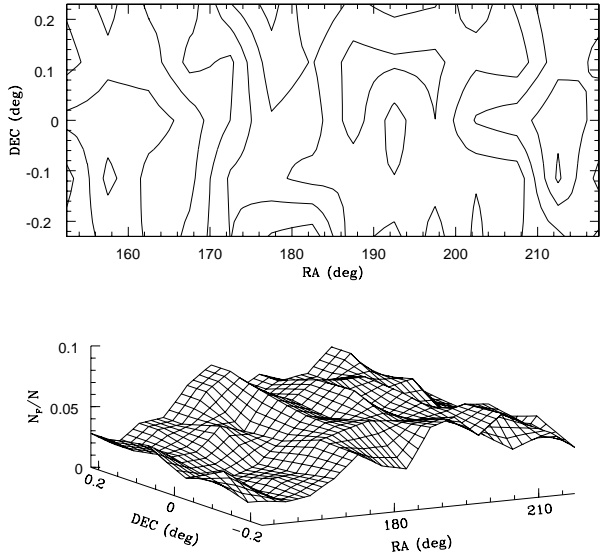,width=\columnwidth}}
\caption{A contour (upper) and surface (lower) plots showing how the F-star distribution ($N_{\rm F}$) compares to the full stellar distribution ($N$). The contours are evenly spaced at $\Delta \frac{N_{F}}{N}$=0.01 intervals between $\frac{N_{F}}{N}$=0.01 and $\frac{N_{F}}{N}$=0.15.}
\label{posplot_rat}
\end{figure}

Further evidence of this feature's reality comes from Fig.~\ref{posplot_rat}, which  compares the projected spatial distributions of
the F-stars and the full stellar samples, by taking the ratio of the
star counts.  Towards the end of the MGC strip there are bins in which
the F-stars are more clustered. In other words, the surface plot in
the lower panel of Fig~\ref{posplot_rat} is not flat, which is what
one might have expected if the F-stars were distributed in the same
manner as the total stellar population. We now investigate other ways
of quantifying this implied F-star clustering.

\subsubsection{Angular correlation function}

An alternative means of quantifying clustering on the sky is through the
well-known angular correlation function, $\omega$($\theta$). This
approach has been used widely to analyse the clustering properties of galaxies
(e.g.~Groth \& Peebles 1977; Shanks et al.~1980; Maddox, Efstathiou \&
Sutherland 1996; Collins, Heydon-Dumbleton \& MacGillivray 1989,
Cabanac, de Lapparent \& Hickson 2000, K\"ummel \& Wagner 2000 and
Couch, Jurcevic \& Boyle 1993). Applications of the angular
correlation function to samples of faint stars have been limited.
Gilmore et al. (1985) derived the two-point correlation
function of their large sample of faint, $B_J < 19.5$, Galactic stars,
derived from photographic plates covering $\la 30$~square degrees at
high Galactic latitude, and showed that it was flat, consistent with a 
`complete absence of clustering'. These authors used this to argue that their 
star-galaxy separation was robust, and that patchy reddening was not important.
They did not quantify what `complete absence' meant in terms of
limiting the existence of substructure. Doidinis \& Beers (1989)
analysed the angular correlation function of 4400 candidate field
horizontal branch stars over 2300 square degrees, finding evidence for
clustering on angular scales of $\la 10^{\prime\prime}$, or physical scales 
of $\la 25$~pc for the characteristic magnitude of the sample. They did not 
consider the clustering properties of more general field stars. 

Here we want to quantify the (absence of?) features seen in the stellar 
angular correlation function. Of the different methods devised for measuring 
$\omega$($\theta$) we adopt the method used in Shanks et al.~(1980) and 
Collins et al. (1989):

\begin{equation}
\omega(\theta)=\frac{N_{ss}}{N_{rr}}-1
\end{equation}

\begin{figure}
\centerline{\psfig{figure=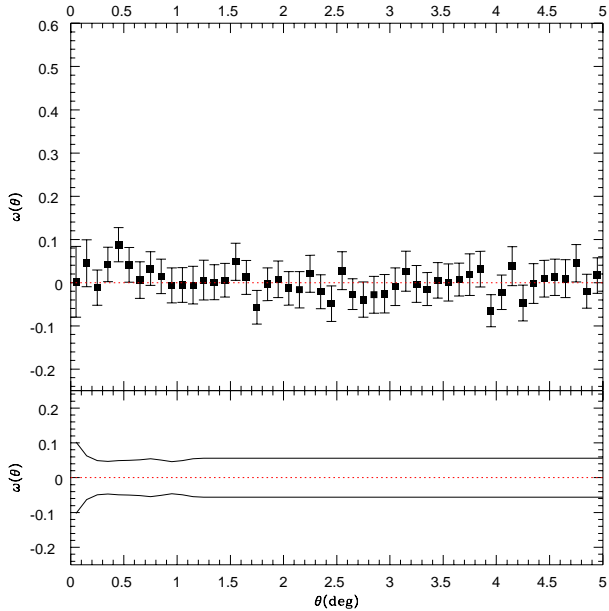,width=\columnwidth}}
\caption{The upper panel shows the angular correlation function,
$\omega$($\theta$), for the bright, $B_{\rm MGC} < 19$, F-stars, as a
function of angular separation, $\theta$. The mean correlation
function is shown by the dotted line and is flat and essentially
zero. The errors shown are $\pm$1$\sigma$ limits about the mean of
each bin. No point is more than 3$\sigma$ deviant. The lower panel
shows the correlation functions of 200 random simulations, taking
account of the overall gradient in stellar number density of the
sample. The solid lines indicate the 1$\sigma$ range of the
simulations correlated about themselves; the fluctuations in the
measured data correlation function are again seen to be at most about
3$\sigma$ away from random.}
\label{wtb}
\end{figure}

\begin{figure}
\centerline{\psfig{figure=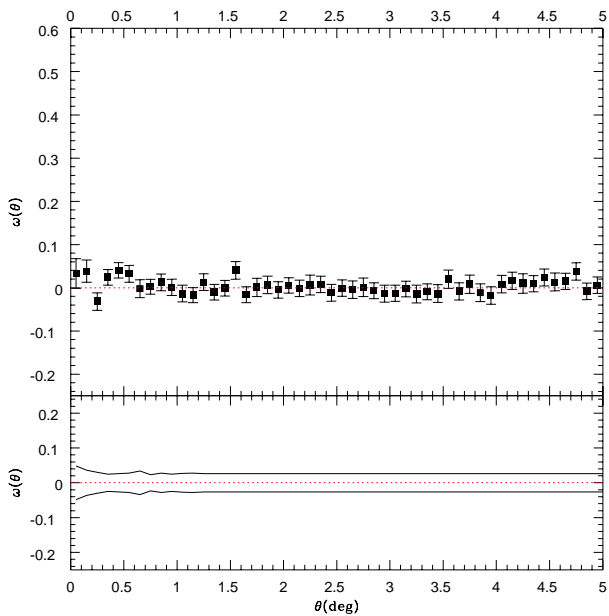,width=\columnwidth}}
\caption{As Figure~\ref{wtb}, but for all the F-stars, with no
magnitude cut. Again the upper panel is the measured angular
correlation function and the lower panel the results of random
simulations. Again no deviations more significant than 3$\sigma$ are
seen.}
\label{wtf}
\end{figure}

\begin{figure}
\centerline{\psfig{figure=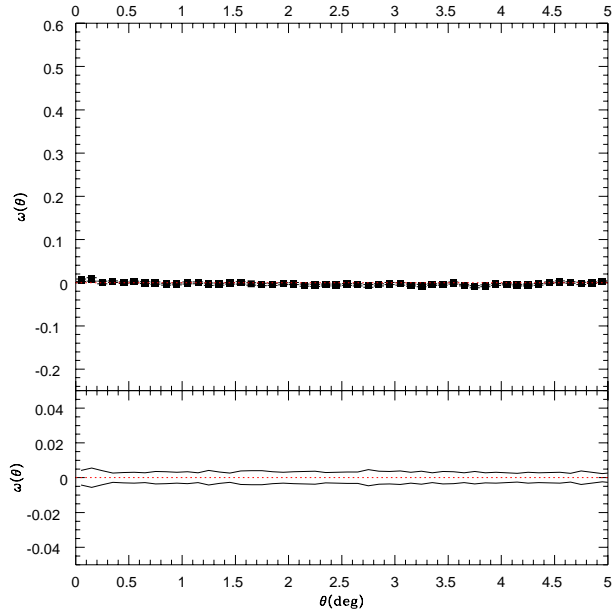,width=\columnwidth}}
\caption{As Figure~\ref{wtb}, but for the full MGC-SDSS matched stellar
sample. Again the upper panel is the measured angular correlation
function and the lower panel the results of random simulations. Again
no deviations more than 3$\sigma$ significant are seen.  }
\label{wtall}
\end{figure}

Here $N_{ss}$ is the number of star-star pairs with separation
$\theta$ and $N_{rr}$ is the number of simulated random-random pairs,
again with separation $\theta$. In practice we follow the same
methodology as Doinidis \& Beers (1989). In what follows, $\omega
(\theta)$ will be derived from the mean random number of pairs
calculated from 200 simulations. The simulations take into account the
Galactic large scale stellar gradients and contain the same number of
simulated stars as found in each sample. We calculate
$\omega$($\theta$) for the range $0< \theta <5$ deg with
$\Delta\theta$=0.1$^{\circ}$. Figs.~\ref{wtb}, \ref{wtf} \&
\ref{wtall} display $\omega$($\theta$) as a function of $\theta$ for a
bright F-star sample (limited brighter than $B_{\rm MGC} = 19$), the
full F-star sample, and the total MGC-SDSS matched stellar
distributions respectively.\footnote{In order to save on computational
time, for the total stellar distribution only 50 random simulations
were used.}

\begin{figure}
\centerline{\psfig{figure=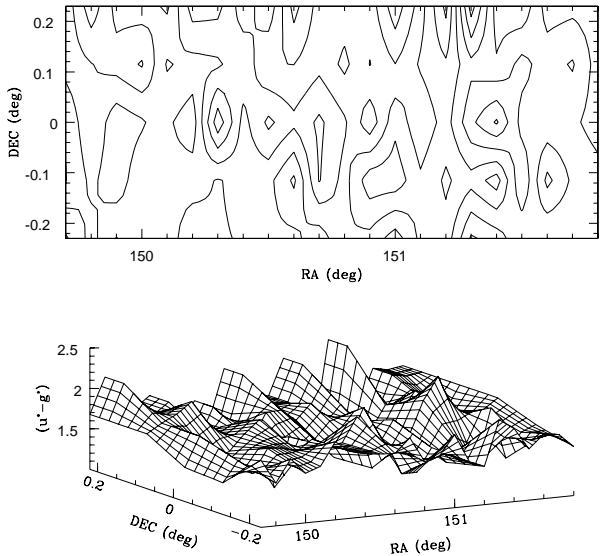,width=\columnwidth}}
\caption{A contour (upper) and surface (lower) plot of the colour
distribution in the first 2 degrees of the MGC strip. The contours
mark the colour range $0.2<(u^{*}-g^{*})<3$ in 0.2 intervals. The
observed fluctuations have narrow colour ranges and are not thus probably
not due to random effects alone (see text). This may reflect reddening
that is patchy on scales of $\sim$1$^{\circ}$. }
\label{cpp}
\end{figure}

The angular correlation functions are essentially flat, with zero mean
angular correlation function, fully consistent with a random
distribution and no excess clustering at greater than the 3-$\sigma$
level of significance. In Fig~\ref{wtf} $\omega(\theta)$ has both a
mean and standard deviation of less than 0.005. The fluctuations
detected may reflect simply statistics, as indicated by the variation
seen in simulated random distributions, or may be in part attributable
to patchy reddening/extinction (see Hewett 1983). To illustrate, the
spatial distribution of stellar colour in bins of size 0.2$^\circ$ by
0.1$^{\circ}$ is shown in Fig.~\ref{cpp} (limited to the first two
degrees of the survey); although there is a mean of only 10 stars in
each RA-DEC bin, the standard deviation in colour ($\sigma_{\overline{
(u^{*}-g{*})}}$=0.2) is significantly smaller than the mean standard 
deviation of colour across all bins (${\overline \sigma}$=0.6). This 
uniformity of the colour indicates that the observed fluctuations are 
not due to random errors alone but that patchy reddening may contribute 
to the fluctuations on scales less than a degree.  

Thus the angular two-point correlation function shows only barely
significant clustering. 

\subsubsection{The Lee 2d statistic}

The angular correlation function contains only low-level suggestion of
any clustering, indeed limiting substructure to very low amplitude.
The counts-in-cells did reveal  substructure, but again at low
significance. When examining just the F-stars and the F-star-all star
ratio (see Figs~\ref{posplot_f} \& ~\ref{magposplot} and
\ref{posplot_rat}) we have found that there are some small
overdensities of faint stars towards the end of the MGC strip, which
if real could be associated with tidal debris from previous pericentre
passages of th Sagittarius dwarf.  We now turn to more innovative
tests for clustering, to see if they can either place more stringent
limitations, or indeed reveal low-level clustering with more
significance.
 
The Lee 2d statistic (Lee 1979) has been used previously mostly in
studies of substructure in clusters of galaxies, and indeed has been
found to be the more sensitive test to the presence of structure
(Rhee, van Haarlem \& Katgert 1991) when compared to the angular
separation test of West, Oemler \& Dekel (1998; 
formally similar to the angular correlation function). A detailed
discussion of the Lee statistic can be found in Fitchett (1988). The Lee 2d 
statistic is essentially a likelihood ratio, with the statistical analysis
analogous to looking for the maximum likelihood split of a two-dimensional 
data set into two similar clumps. 

The Lee statistic is calculated by projecting the data perpendicularly on to 
a line. At each point along this line a measure of the clumpiness of the data 
is found and a maximum determined, L($\theta$). The clumpiness is determined 
at each point/partition along the line by measuring the deviation away from 
the mean (or Gaussian distribution) of all the points to the left and right 
of the partition. This is then compared to the deviation away from the mean 
of the total data set. In this way any underling large-scale structure 
gradient is ``flat fielded'' out as part of the analysis process. The line is then rotated by some small 
amount, $\Delta$ $\theta$, and the process is repeated. From this we can plot 
L($\theta$) against $\theta$, the maximum of which, L($\theta$)$_{\rm max}$, 
indicates the position angle of a line partitioning the data into two clumps. 
The Lee 2d statistic has been used to look for substructure within galaxies, 
clusters of galaxies (Fitchett \& Webster 1987; Rhee et al. 1991) and in 
simulations of clusters (Crone, Evrard \& Richstone 1996). We now apply it to 
our high-latitude stellar data set.

\begin{figure}
\centerline{\psfig{figure=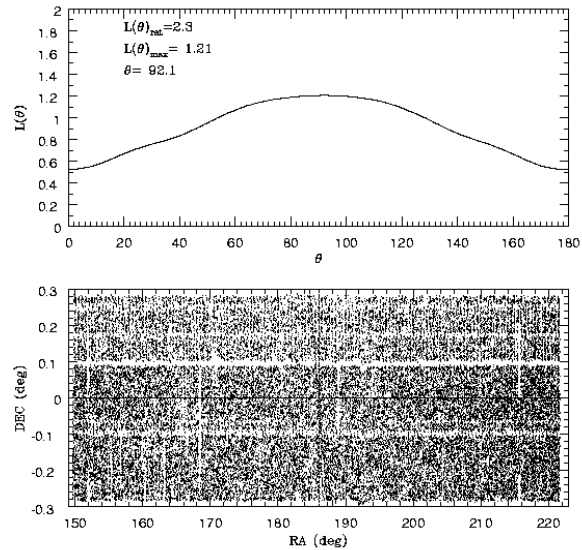,width=\columnwidth}}
\caption{Upper: A plot of L($\theta$)$_{\rm max}$ for the high
latitude stellar sample. Lower: The full MGC-SDSS stellar sample on
the sky.}
\label{lee2d}
\end{figure}

We first apply this test to the (projected) angular distribution on
 the sky of the entire MGC-SDSS matched stellar data set. The data are
 shown in the lower panel of Fig.~\ref{lee2d} and the values of the
 Lee statistic L($\theta$) for this sample are shown in the upper
 panel. The position angle, $\theta_{\rm max}$, of the maximum of the
 Lee statistic is indicated in the lower panel.

\begin{figure}
\centerline{\psfig{figure=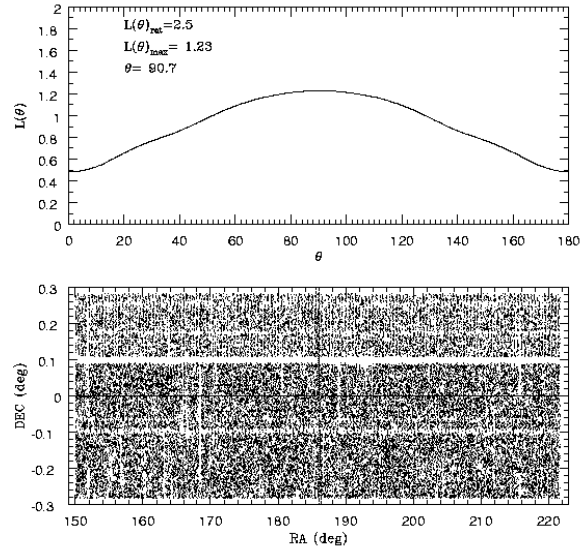,width=\columnwidth}}
\caption{Upper: A plot of L($\theta$)$_{\rm max}$ for the simulated high latitude stellar sample. Lower: The full simulated stellar sample.}
\label{lee2d_sim}
\end{figure}

The L($\theta$) curve shows a rather broad peak, indicating only a
low-significance detection of an over-density; a stronger detection
would result in a more obvious peak. The value of the angle at which 
the Lee statistic has its maximum, $\theta_{\rm max}=92.1$, indicates that the 
overdensity is located towards the top right of the lower plot, again at 
high values of RA (as in the earlier statistical tests). The significance of 
this detection can be quantified by the analysis of randomly distributed 
datasets. Thus we fill the MGC survey area with the same number of stars as 
found in the high latitude sample, but randomly distributed (remember that any
large-scale gradient is removed by the analysis technique). 
Fig.~\ref{lee2d_sim} shows the simulated dataset, as well as the corresponding 
L($\theta$) plot. The angle at which L($\theta$)$_{\rm max}$ occurs is
now 90.7$^{\circ}$, the slight offset from the perpendicular being
understandable as due to the uneven distribution of MGC exclusion
regions. 

	We note that for the simulated data the $L_{\rm rat}$ statistic is 
higher, indicating a more significant detection (of the MGC holes), than in 
the real data. This can be explained by noting that in the simulations there 
was no minimum star-star separation used. Therefore the simulated stellar 
distribution will be slightly smoother than the real data, making the effects 
of the MGC holes more apparent in the simulation. This implies that any result
 from the Lee 2d test will provide only a rough estimate on the size of the 
clustering signal. The lack of a minimum star-star separation in the simulated
 data should not have affected the $\omega (\theta)$ test as it is only 
dependent on the number of stars at a given radius and not their distribution 
about a partition.

Taking the ratio of L($\theta$)$_{rat}$ $\left(=\frac{L(\theta_{\rm max})}
{L(\theta_{\rm min})} \right)$ for the simulated and real data we find that 
the over-density seen in the MGC region is detected at 8 per cent 
$\left(={100\times\bmod\left(1-\frac{L(\theta)_{rat_{MGC}}}{L(\theta)_{rat_{MGC_{sim}}}}\right)}\right)$ above a smooth stellar distribution.

To summarise the Lee 2d scheme, a line at some angle, $\theta$, is
drawn dividing the data in half. Every star is then projected onto
that line, defining a partition point for each star. The
clumpiness/asymmetry parameter L$(\theta)$ is then calculated for all
stars/partitions along the line and the largest value of L$(\theta)$
is found for that line/angle. The MGC RA and DEC coordinates of the
star whose partition gave that value of L$(\theta)$ is then used as
the coordinates of the over-density. By doing this for the entire stellar sample from the MGC,  we find a weak
clustering signal at RA=$216.5^{\circ}$ and DEC=$0^{\circ}$,
consistent with visual inspection of the distribution of the
stars. This is consistent with the counts-in-cells analysis, and again if real 
can be a combination of  A-stars from  the distant, most recent, tidal debris from the Sagittarius dwarf (cf.~Yanny et al.~2000; Vivas et al.~2001; Ibata et al.~2001), and F-stars from an older, more nearby, tidal stream. 

\begin{figure}
\centerline{\psfig{figure=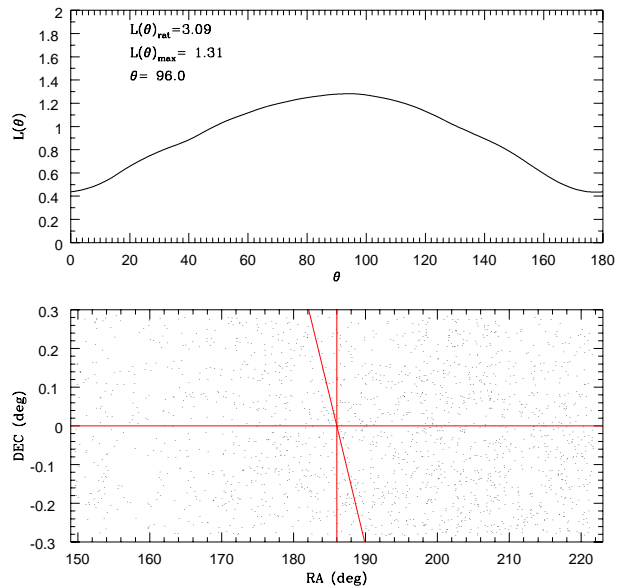,width=\columnwidth}}
\caption{The Lee 2d test for the F-star sample. The upper and lower panels 
are the same as in Fig.~\ref{lee2d}.}
\label{lee2d_F}
\end{figure}

\begin{figure}
\centerline{\psfig{figure=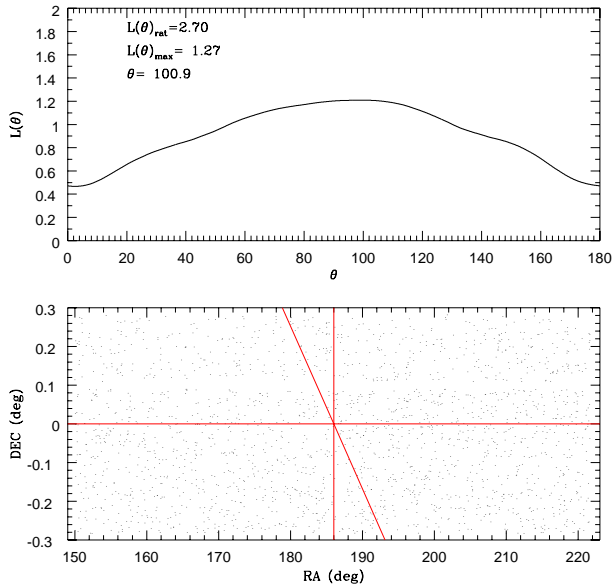,width=\columnwidth}}
\caption{The Lee 2d test for the simulated F-star sample. The upper and lower 
panels are the same as in Fig.~\ref{lee2d_sim}.}
\label{lee2d_simF}
\end{figure}

Having shown that the Lee 2d statistic is sensitive to substructure,
we now investigate the signal for the inner halo by using only the
brighter F-stars, in the magnitude range $17.5 <B_{\rm MGC}<19$.
Figs.~\ref{lee2d_F} and \ref{lee2d_simF} show the results of the Lee
2d test for the real and simulated (random distribution) F-star
population.  We find a slightly stronger clustering signal by
isolating F-stars, with an amplitude of 14 per cent deviation from a
random distribution, with the clustering now placed at
RA=$205^{\circ}$ and DEC=$0^{\circ}$. Again this places it within the
2-D coordinate space covered by a Sagittarius dwarf tidal stream, with
the inferred distances (if F-stars and not A-star contamination)
pointing to debris from an older pericentre passage (Johnston et
al.~1999).  We find no other evidence of clustering.

\subsection{Halo Flattening Revisited}

\begin{figure}
\centerline{\psfig{figure=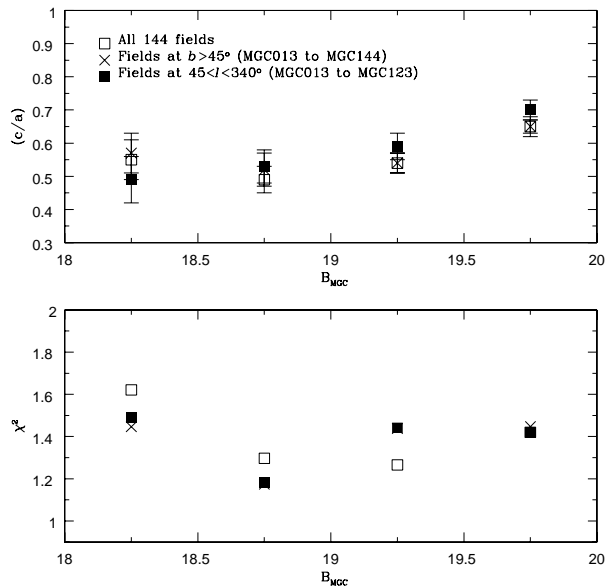,width=\columnwidth}}
\caption{Here we show the effect on the favored axial ratio of the
stellar halo of removing the low ($b<45^{\circ}$) Galactic latitude
and high ($l>340^{\circ}$) Galactic longitude fields from the
$\chi^{2}$ fitting.}
\label{chisqu}
\end{figure}

Although there is no strong detection of structure within the Galactic
halo sampled, we decided, to be conservative, to re-fit the star-count
models to a restricted sub-sample. We now fit to star count data at
$b>45^{\circ}$ and for data in the range $b>45^{\circ}$ and
$l<340^{\circ}$ (see Tables~\ref{data1} \& \ref{data2}). This was done
to minimize the contributions from the thin/thick disks and the
thin/thick disks and bulge respectively. Note that this longitude cut
also serves to remove any effects of the Sagittarius Dwarf tidal
stream. Fig.~\ref{chisqu} shows the effect of removing the low
latitude and high longitude fields from the $\chi^{2}$ fitting; a
stellar halo with an axial ratio of ${\rm (c/a)}$=0.52$\pm$0.05 now
provides the best fit to the data.  The sample is dominated by stars
within $\sim 10$~kpc of the Sun, and this axial ratio thus applies to
the inner halo.

\begin{figure}
\centerline{\psfig{figure=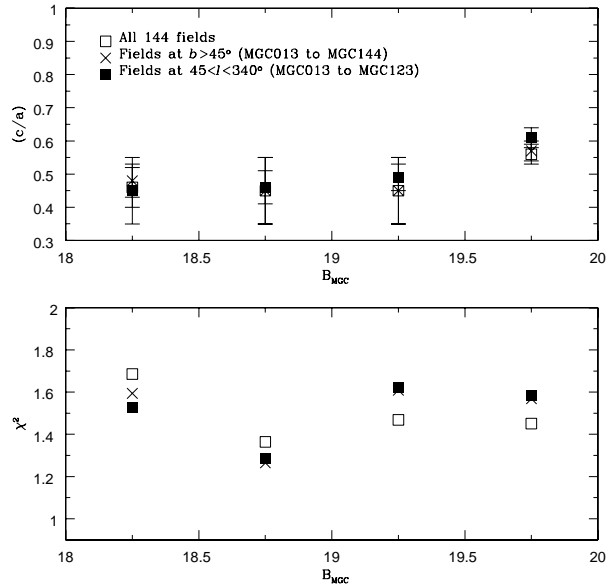,width=\columnwidth}}
\caption{Here we show the effects of using a potentially cleaner
sample, one consisting of only MGC-SDSS matched stars with
$(u^{*}-g^{*})>0.6$.}
\label{chisqu2}
\end{figure}

A further refined sample contains only MGC-SDSS matched stars with 
$(u^{*}-g^{*})>0.6$. Although there are 2418 fewer stars in this subsample 
than in the full stellar sample 1021 of the rejected stars having 
$(u^{*}-g^{*})<0.6$, this sub-sample should now be free from any spurious 
effect that may have been induced by the possible inclusion of QSOs 
(previously identified as the fuzzy patch in Fig.~\ref{colcol}).

Table~\ref{data3} shows the best fit halo axial ratio in each
magnitude bin for the sample with Galactic latitude and longitude cuts, 
and Fig.~\ref{chisqu2} compares the results from this reduced sample to the 
original one. In fact the $\chi^{2}$ values and overall shape of the 
``axial-profile'' remain more or less the same. It is only the value of
 the halo axial-ratio that has changed, with the reduced sample preferring a
slightly more flattened halo, with ${\rm (c/a)=0.45\pm0.1}$. Note that 
as the models are only generated within the range $0.45\le {\rm (c/a)} \le 
0.8$, any `best-fit' model with an axial ratio less than 0.45 is spurious 
and simply represents a lack of stars within the magnitude bin and $b$ and 
$l$ range being tested.

The star-count models are then fitted to the full MGC stellar sample, to the 
MGC-SDSS sample (i.e. no colour cuts) and to the F-star sample in the magnitude range 
$18<B_{\rm MGC}<19$, both with and without the $b$ and $l$ cuts. In doing 
this we find that the best fit comes from the full MGC stellar sample with 
both the $b$ and $l$ cuts and gives an axial ratio of (c/a)=0.61 $\pm$0.02. 
However, the axial ratio with the smallest error is given by the F-star 
sample, again with both the $b$ and $l$ cuts, (c/a)=0.56$\pm$0.01.  This is again sampling the inner halo.

\begin{figure}
\centerline{\psfig{figure=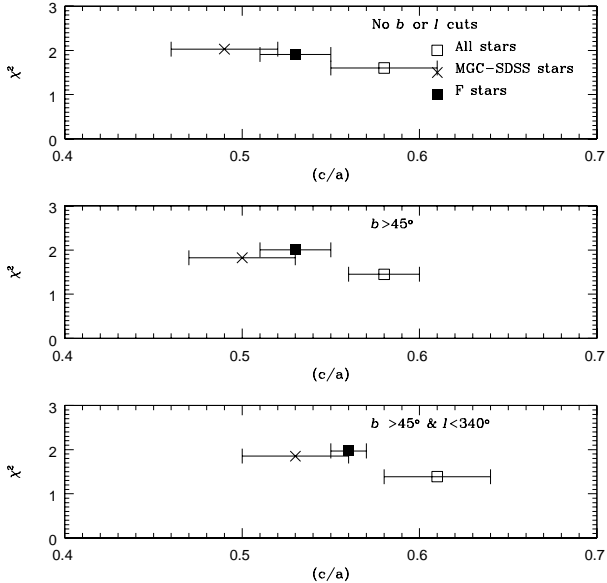,width=\columnwidth}}
\caption{A graphical summary of all the best fit star-count models for
each stellar sample.}
\label{sum_ca}
\end{figure}

In comparing our results with previous work we find that our final
value is in very good agreement with the pioneering work of Kinman,
Wirtanen \& Janes (1965) who reported (c/a) = 0.6 for the stellar halo
interior to the Sun, based on a small (less than 100) sample of bright
RR Lyrae stars. Indeed most star-count analyses of the inner halo give
a flattening of this order (e.g.~Hartwick 1987; Wyse \& Gilmore 1989;
Larsen \& Humphreys 1994, 2003; Siegel et al.~2002), consistent with
anisotropic velocity dispersions (Wyse \& Gilmore 1989; Chiba \& Beers
2000).  The outer stellar halo may be rounder (e.g.~Hartwick 1987;
Preston, Shectman \& Beers 1991), perhaps reflecting a differing
importance of dissipation during formation of the inner and outer halo
(cf.~Norris 1994; Chiba \& Beers 2001).  A detailed investigation of
the variation of flattening with distance is beyond the scope of the
limited data given here, but well within the scope of the final SDSS
dataset. However we do note that our more local F-star analysis yields
a marginally significantly lower (c/a) than the full stellar
population. Furthermore we note that within our analysis we see a
marginal increase in (c/a) with magnitude (distance), (Fig.~3, 20, \&
21), however this could also be explained by contamination from the
QSO and/or galaxy population at the star-galaxy separation limit.

\section{Conclusions}

	We have used a sample of 42457 stars from the MGC-Bright
photometric catalogue, most of which ($\sim 97$ per cent) have
SDSS-EDR counterparts, to investigate the structure of the stellar
halo of the Milky Way. The MGC-SDSS matched objects were used to
define colour selection limits allowing us to isolate a sample of
F-stars.  Our major results are the first quantification of the
clustering in coordinate space in the inner halo, plus a new estimate
of the large-scale flattening of the inner halo.

	We conclude that the stellar halo of the Milky Way is
significantly flattened, with an axial ratio of (c/a)=0.56$\pm$0.01
(within $R \le 10$ kpc). While this result is in line with previous
determinations, it is more robust, having been derived from samples
cleaned of substructure and with contamination from quasars and other
stellar components minimised.

We find weak evidence for substructure in the brighter F-stars,
tentatively identified with tidal debris from the Sagittarius dwarf.
We find no other substructure.  This essentially null result implies
that recent accretion of stellar systems is not important in the bulk
of the stellar halo. Quantification of limits on the disruption rate
of putative satellites needs to be investigated, but is beyond the
scope of the present paper. However these firm limits of observable
clustering need to be included in future hierarchical clustering
models.

We thank Gerry Gilmore for updated star-count models. RFGW
acknowledges receipt of a Visiting Fellowship from PPARC and thanks
all for a stimulating and pleasant environment during her sabbatical.

\section*{Tables}
\begin{table*}
\caption{The best fit model in each magnitude bin using all the data.}
\begin{tabular}{lll} \hline
$B_{\rm MGC}$ & (c/a) & $\chi^{2}$ \\ \hline 
 18.0 $<B_{\rm MGC}<$ 18.5 & 0.55 $\pm$ 0.06 & 1.62\\
 18.5 $<B_{\rm MGC}<$ 19.0 & 0.49 $\pm$ 0.04 & 1.30\\ 
 19.0 $<B_{\rm MGC}<$ 19.5 & 0.54 $\pm$ 0.03 & 1.27\\
 19.5 $<B_{\rm MGC}<$ 20.0 & 0.65 $\pm$ 0.03 & 1.42\\ \hline
\end{tabular}
\label{alldata}
\end{table*}

\begin{table*}
\caption{The best fit model in each magnitude bin using all data at $b>45^{\circ}$.}
\begin{tabular}{lll} \hline
$B_{\rm MGC}$ & (c/a) & $\chi^{2}$ \\ \hline 
 18.0 $<B_{\rm MGC}<$ 18.5 & 0.57 $\pm$ 0.06 & 1.45\\
 18.5 $<B_{\rm MGC}<$ 19.0 & 0.52 $\pm$ 0.05 & 1.17\\ 
 19.0 $<B_{\rm MGC}<$ 19.5 & 0.54 $\pm$ 0.04 & 1.44\\
 19.5 $<B_{\rm MGC}<$ 20.0 & 0.65 $\pm$ 0.04 & 1.45\\ \hline
\end{tabular}
\label{data1}
\end{table*}

\begin{table*}
\caption{The best model in each magnitude bin in using all data at $b>45^{\circ}$ \& $l<$$340^{\circ}$.}
\begin{tabular}{lll} \hline
$B_{\rm MGC}$ & (c/a) & $\chi^{2}$ \\ \hline 
 18.0 $<B_{\rm MGC}<$ 18.5 & 0.49 $\pm$ 0.08 & 1.49 \\
 18.5 $<B_{\rm MGC}<$ 19.0 & 0.53 $\pm$ 0.06 & 1.18 \\ 
 19.0 $<B_{\rm MGC}<$ 19.5 & 0.59 $\pm$ 0.04 & 1.44 \\
 19.5 $<B_{\rm MGC}<$ 20.0 & 0.70 $\pm$ 0.03 & 1.45 \\ \hline
\end{tabular}
\label{data2}
\end{table*}

\begin{table*}
\caption{The best model in each magnitude bin using colour selected stars, with $(u^{*}-g^{*})>0.6$, at $b>$$45^{\circ}$ \& $l<340^{\circ}$.}
\begin{tabular}{lll} \hline
$B_{\rm MGC}$ & (c/a) & $\chi^{2}$ \\ \hline 
 18.0 $<B_{\rm MGC}<$ 18.5 & 0.45 $\pm$ 0.10 & 1.53 \\
 18.5 $<B_{\rm MGC}<$ 19.0 & 0.46 $\pm$ 0.10 & 1.28 \\ 
 19.0 $<B_{\rm MGC}<$ 19.5 & 0.49 $\pm$ 0.06 & 1.62 \\
 19.5 $<B_{\rm MGC}<$ 20.0 & 0.61 $\pm$ 0.04 & 1.58 \\ \hline
\end{tabular}
\label{data3}
\end{table*}

\label{lastpage}

\end{document}